# Creative Autonomy Through Salience and Multidominance in Interactive Music Systems: Evaluating an Implementation

Fabio Paolizzo[a,c] and Colin G. Johnson[b]

[a]Department of Cognitive Sciences, University of California, Irvine, Irvine, USA

[b]School of Computing, University of Kent, Canterbury, UK

[c]Department of Electronic Engineering, University of Rome Tor Vergata, Rome, Italy

**Abstract**

*Interactive music systems always exhibit some autonomy in the creative process. The capacity to generate novel material while retaining mutuality to the interaction is proposed here as the bare minimum for creative autonomy in such systems. Video Interactive VST Orchestra is a system incorporating an adaptive technique based both on the concept of salience as a means for retaining mutuality to the interplay and on multidominance in the adaptive generation process as a means for introducing novelty. We call this property reflexive multidominance. A case study providing evidence of such creative autonomy in VIVO is presented.*

## Introduction

Interactive music systems (IMSs) have both extended existing approaches to music making and introduced entirely novel ways for musical creativity. Several types of such systems and frameworks have shaped the literature of the recent past (Drummond, 2009). A most well-known, yet evergreen paradigm for IMSs suggest that these can be defined within a continuum between their capacity to extend the creativity of the user and the resemblance of capacities which are typical of human players (Rowe, 1992). More recently, in relation to the behaviour that a computationally creative agent can exhibit in generating an outcome, characteristics such as novelty, value and intentionality are identified as determining whether such an agent can be categorized either as a tool for creative support, co-creation or as fully autonomous (Ventura, 2017). These definitions capture IMSs at a wide range but, interestingly, they all leverage on the concept of creativity as a function of autonomy. That is, any IMS exhibits a capacity to operate at a creative level, which depends on how autonomous the system is.

Not all types of autonomy can allow a machine to interact as a human could do. For example, a machine musician could be producing outcome that is very coherent within its own scheme, but which does not necessarily produce anything meaningful to a human interactor. Vice versa, a machine could not be able to distinguish which information is particularly meaningful to the human agent. Although the level of complexity in the design of IMSs varies greatly, autonomy always remains a central property of any IMS because their creative capacity is a function of that autonomy, as mentioned. Specific features define how autonomy van allow an IMS to be regarded as creative. We are interested here in investigating the bare minimum of properties that defines creative autonomy in an IMS. We hypothesize that it is not necessary for IMSs to be capable of understanding neither the text nor the context of interaction for such systems to exhibit creativity.

In the next section, we investigate further the concept of creative autonomy and provide reference to some of the most well-known approaches enabling that in IMSs. In section 3, we present an IMS that implements creative autonomy at its bare minimum of features. In section 4, we propose a case study adopting that system and supporting our thesis.

1. **Autonomy in IMSs**

In order to develop a working definition for IMSs that can explain the phenomenon of creative autonomy rather than its ontology, we shall recall that a system may be considered as autonomous when its behaviour resembles the intentionality typical of a human player, as mentioned. Here, resemblance means that the system could 'fool someone into thinking it was human' and/or 'has/suggests a similar level of intentionality as a human'. This concept extends slightly a definition that we mentioned in the introduction (Rowe, 1992). Specifically, the concept links into debates from the computational creativity literature as to whether computers can be creative in ways that are as creative as humans (Boden, 1998; Colton, 2008), whilst clearly being computers and exercising their creativity in a way that is native to computers (Dartnall, 2013; Kantolaso & Riihiaho, 2018). Although autonomy is typically synonymous with complete independence, here we start investigating creative autonomy from a definition of autonomy as a form of self-determination that is not necessarily free from the influence of some external information fed to the system (Bown & Martin, 2012). From this perspective, a distinction can exist between an autonomous IMS with a seemingly random behaviour, and one that is controllable or influenceable by the user. Thus, we are here framing system autonomy in terms of a control that is exercised by the user, rather than merely being implemented by design. Notably, the exercise of control does not imply that an agent is aware of being in control. This is a well-known phenomenon in cognitive sciences (Wyer Jr. & Srull, 1994; Tsakiris & De Preester, 2018). For example, in a musical interplay with an IMS, a user may perceive a system that is too autonomous or too predictable as either being an unengaging or over-improvisatory partner, regardless of the actual system properties (Bown & Martin 2012; Ornes, 2019; Yu, 2019).

In the present paper, we suggest that a player can maintain control over a specific set of parameters, while the system exhibits dynamical and unpredictable sonic behaviour. This is a common feature in various IMSs, even for simple interactions 'altering the relation the system has to itself' (Sanfilippo, 2012; 2015). Accordingly, a player could consider an IMS to be autonomous when capable of dealing with external or internal inputs in ways that change its behaviour. This understanding allows us here to identify a property that a system must have for a system to be perceived as autonomous. Because an IMS operates interactively, rather than automatically, we expect the generation in an IMS to allow for the retention of some mutuality to the context in terms of user's action and perception, as it occurs alongside an interplay or co-invention. The capacity to exchange information with the context for informing the text is a determining factor in creative autonomy. In the next section, we reference a few IMS that are well-known for allowing the user to recognise mutuality as a property of the system.

*Mutual listening*

A well-known approach to the design of computer programs whose behaviour mimics that of human interplayers in a musical improvisation consists of using algorithms that monitor the improvisation and use the information gathered to generate new and contextually relevant material. A listener understands the computer outcome in terms of a response to a musical gesture from the human player. Early examples of this are GenJam

(Biles, 1999) and MusicBlox (Gartland-Jones, 2003), which use interaction and interactive genetic algorithms to define the quality of the contextual fit between the computer-mutated musical fragment and the human performer's contribution. The formalisation deriving from a definition of the initial population and the use of interaction rules mitigate the capacity for novelty because the (user-dependent) decision-making process is subjective and unilateral.

*Multidominance*

Other mutual listening works (Chadabe, 1984; Perkis, 1999; Brown & Bischoff, 2002) use the combined behaviour of software and human agents to determine overall system complexity. In terms of system autonomy, this is an improvement over unilateral human-to-machine interactions, as multiple input (musical) gestures are re-interpreted into a complex musical output. Such systems denote a form of shared control. However, only response-response interactions can be determined, as the software agents do not exhibit a capacity for **multidominance**, a term borrowed from Douglas (1991), meaning a form of interaction in which all participants contribute primary material. As such, these systems cannot lead the musical direction of the performance because the primary generator of novel material is only the human performer. Multidominance as a system property is also a trait of authenticity and authorship in autonomy, and one of the first systems capable of such style-independent response is Voyager (Lewis, 2000). Voyager carries out sonic behaviour grouping by imitating, opposing, or ignoring the performer's musical dynamic. The system then processes outcomes and reconfigures any algorithm involved in the grouping with 'no built-in hierarchy of human leader/computer follower' (Lewis, 2000, pp. 36).

*Modelling knowledge*

A computational music model can be achieved by segmenting music sequences in a corpus and analysing those segments for common elements of style. These elements can then be used to recombine the segments into new works (e.g., Cope, 2010; 2016), also using dictionary-based machine-learning models (Dubnov, 2003), or to impose stylistic constraints over different material (Pachet, 2016). Similarly, by operating within a machine-learning scheme, music expectation can be modelled (Weng, 2010), as well as mutual listening. For example, OMax learns 'in real-time by listening to an acoustic musician and extracting symbolic units from this stream. It then builds a sequence model on these units constituting an internal knowledge' (Lévy, Bloch and Assayag, 2012: 1). This type of algorithms can navigate the model and recombine the musician's discourse, who is exposed to a form of stylistic reinjection: the system constantly confronts the player with 'a reinterpreted version of his own playing' (Lévy et al., 2012, pp. 1). While these approaches may allow a system to operate in autonomously creative ways, their complexity exceed the bare minimum which we seek in the present article. To this purpose, we shall recall that, simply, '[s]trong interactivity depends on instigation [by the system] and surprise [by the human performer], as well as response' (Blackwell & Young, 2005).

## 2. Video Interactive VST Orchestra

Video Interactive VST Orchestra (VIVO) is an IMS that we define as **wakeful** (Paolizzo, 2013). Here, the term wakeful consists in the idea that computers and human beings can interact via an evolving language, which is expressed in VIVO as a form of simulating intentionality in music generation derived from a process of adaptation to the user's interaction.

VIVO implements theories from previous research on music and interaction (Paolizzo, 2006), which investigates the relation between the human capacities for interpreting and interacting with music. In a typical scenario (Figure 1), the user controls a sound source (i.e., a musical instrument), which sends an audio signal to VIVO for sound processing. In this scenario, VIVO observes the user's movement by the means of a camera connected to the system and analyses the information to control the sound processing automatically. In this specific regard, the system resembles VNS (Rokeby, 2010), which also implements an approach for gesture/video-to-sound types of mapping. The key components of VIVO which differentiate it from similar systems are highlighted in the next sub-section. In the scenario being discussed here, the audience can hear both the original and unprocessed sound source together with the sound that VIVO generates. In other scenarios, the sound or the information that controls the processing could be derived from different types of sources, such as a video or an external device connected to the system (e.g. haptic, text-based, etc.). In any scenario, the system simply analyses the user's interaction and uses that information to control the processing of the audio signal to generate a subsequent musical output.

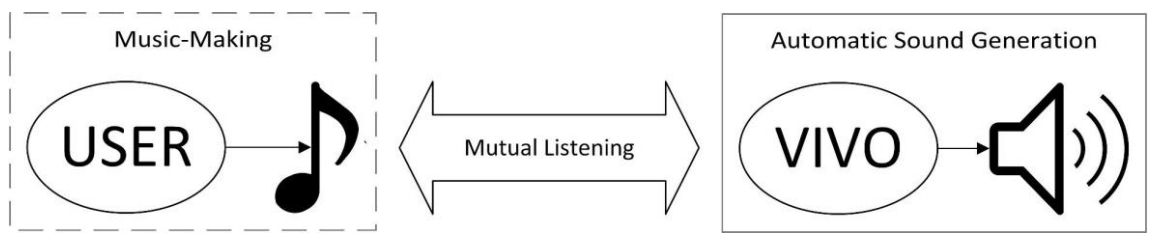

Figure 1. Typical interaction model for a VIVO/user instance.

*Design of VIVO*

VIVO is an open source computer program developed in MAX/MSP (Cycling 74, 2017), which is capable of real-time audio processing and sound synthesis by loading and using external audio plug-ins (VST, VSTi, DirectX, AU) in the program. Figure 2 shows an overview of the system architecture. The system is comprised of different software components: (a) an adaptive video tracking module, (b) a variance-based threshold trigger, (c) an editor for stochastic scores, (d) a single graphical user interface for every audio plug-in loaded, (e) a dynamic host for automating the activation and control of the audio plug-ins, and (f) network and web components to send and receive external data for extended configurations.

In the present study, we focus on the influence and implications of (b) the variance-based threshold trigger on the interplay, through the action of (e) the dynamic host, when receiving data from (a) the adaptive video tracking module. The software components (b) and (a) allow the simultaneous detection and mapping both of the Quantity of Motion (QoM) — the amount of movement in a video stream (Camurri et al., 2003; Paolizzo, 2013) — and of the Salience of Action (SoA) — herein defined as the variance, S, either of the QoM or of the amount of change directed by the user on the mapped parameters. The variance-based threshold trigger determines the SoA, which influences the constraints imposed by (c) the editor, is shown by (d) the user interface, and instructs the automatic sound generation in (e) the audio processing sequence of the dynamic host.

Multidominance is implemented in VIVO by a cross-modal design, through the combined use of (a) the adaptive video tracking module and (e) the dynamic host. SoA builds on the concept of **salience**, implemented through (b). Salience is a factor informing the human interplayer about the potential effects of his/her musical actions on the interplay, similarly to a 'vested interest' influencing the subject's self-efficiency (Crano, 1995). As shown in our previous research (Bowman et al., 2012), a computational detection of salience within a data stream representing aspects of the interaction process can be used to manifest the potential for an interplayer to act. Implementing salience allows reflecting the users' musical interpretation in the computer outcome; salience is the principle by which mutuality to the interplay is implemented in VIVO. Creative autonomy is sought via salient sonic changes that are synchronic to the user's interaction with the system, which in turns affords a dynamic amount of control to the user. In terms of musical interpretation, salience is therefore a variable describing the communication between sign producers. The design of VIVO introduces salience-based techniques capturing the user's intentionality and adopting that information to mould the automatic generation accordingly. Human and software agents can identify salience in the information they are generating. In this process of meaning attribution, the changes that a salience-based generation produces create an expectation for meaning to be found in the outcome, both for the user and for the audience. This process provides a basis for identifying the automatic sound generation as novel music via the user's reflexive interaction. Such a generation framework offers a creation capacity that is open-ended, because third-parties external audio plug-ins can be loaded into the system. Implications of the system architecture are discussed throughout the rest of the present article.

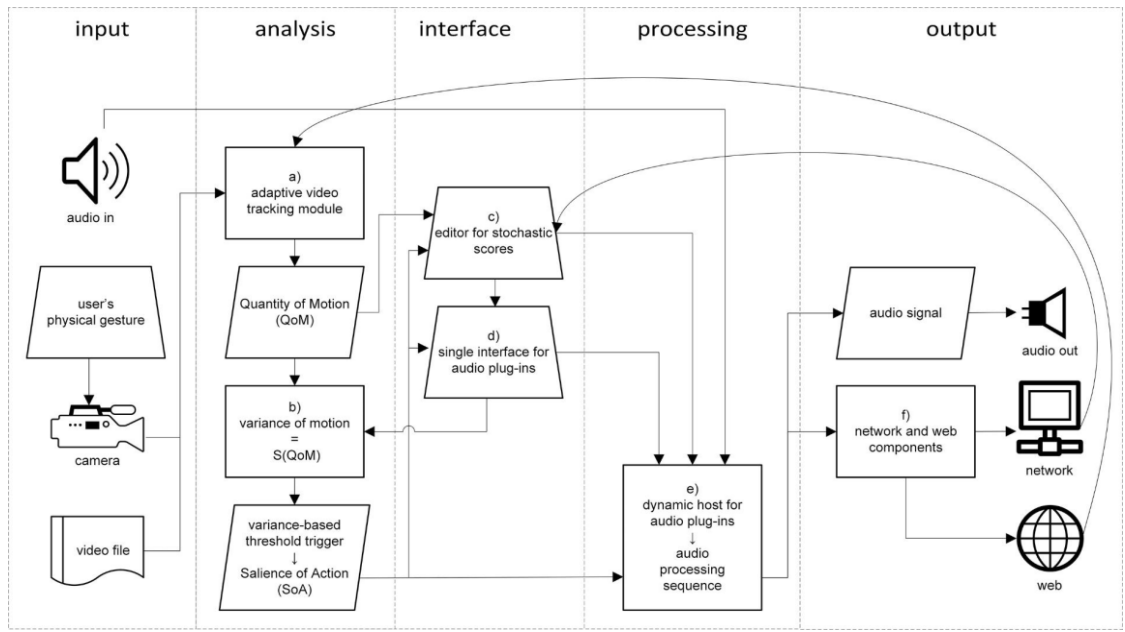

Figure 2. System architecture of VIVO.

In the typical interaction model of a VIVO/user instance (Figure 3), a user engaging in music-making also enacts gestures with a musical intention (i.e., physical gestures on a musical instrument). VIVO extract salient information from these gestures in order to generate a sonic outcome. The user is thus caught in an action-reaction loop of self-reflection (Paolizzo, 2010). In section 2 of the present article, we have discussed some IMSs relying on mutual listening between player and instrument. Such systems operate in terms of salience, implicitly. By detecting and using salience for sound generation, an IMS can influence the musical conduct much as a human interplayer could. Salience-based systems generally derive the data for the generation from the music played by the human interplayer, as presented in Figure 1. It is important to note that the explicit use of salience can allow drawing effectively from non-musical information which is sequentially structured (i.e., language, visual sequencing, motor planning), when such extra-musical information retains some coherence to the interplay (i.e., QoM, SoA). In the next section, we discuss the use of saliency deriving from non-musical but relevant information as a form of multidominance that introduces novelty in the generation while also retaining mutuality to the interplay.

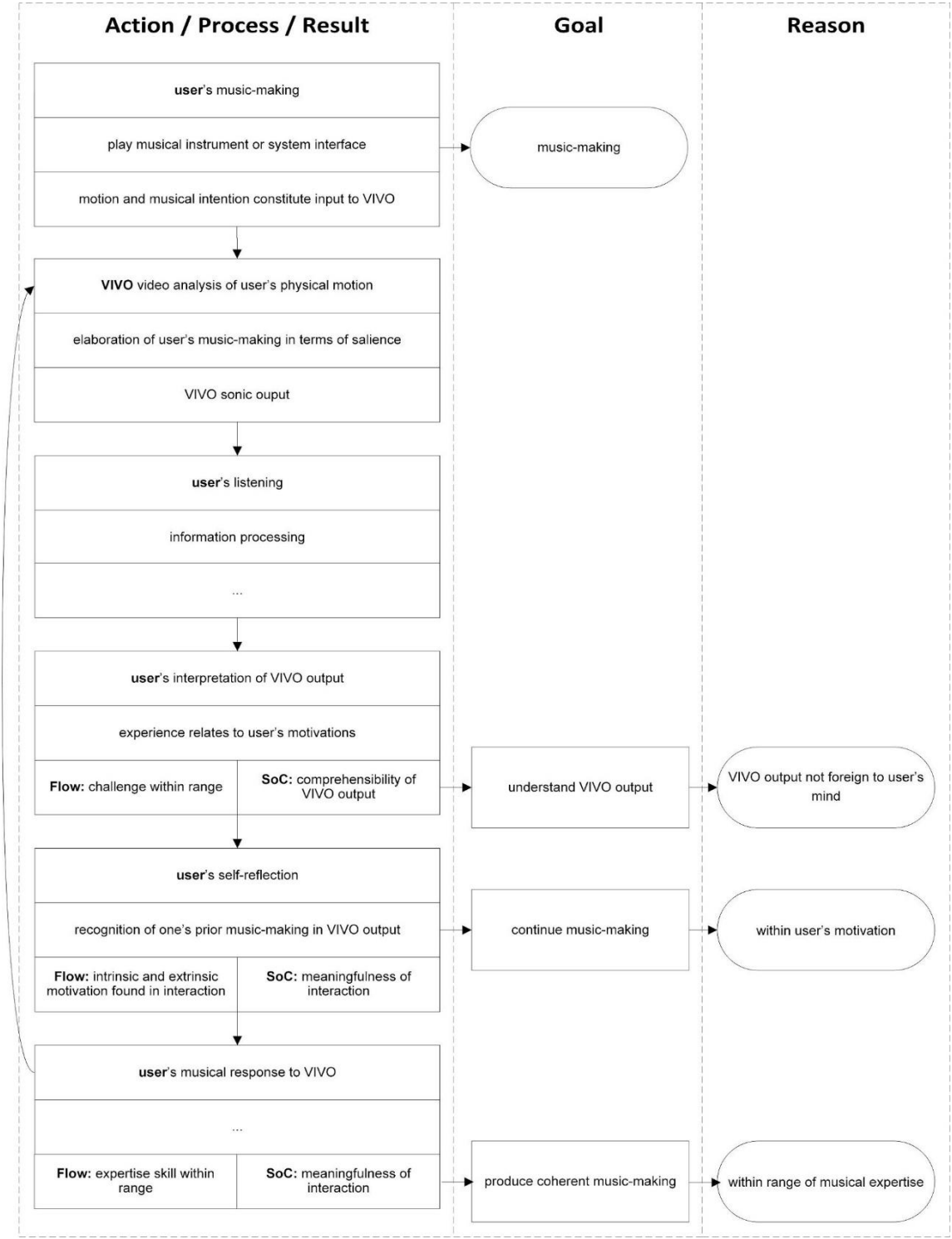

Figure 3. Interaction diagram of a VIVO/user instance.

## 3. Overview of pilot studies

The present research has included pilot studies in which VIVO was used for music-making within a variety of scenarios (Table 1). The purpose of these studies was to test the functionality of the system and to highlight implementation strategies that could maximise the perception of creative autonomy within an action/perception feedback loop for both the user and the audience. In order to provide a framework for the case study that we present in section 4, we introduce and discuss some of the theoretical background underlying the pilot studies. This framework incorporates the concepts already discussed in the present article, such as multidominance and mutuality.

Table 1. Brief description of the pilot studies (selection).

| | |
|---|---|
| *VIVOtube* | An audio/video installation and a dedicated internet page hosting the installation simultaneously. It allows on-site and internet users to search and select videos from YouTube and load it in VIVO for sound processing. The generated soundtrack resulted from a mapping between the sound processing to the motion of the images detected in the video. |
| *Studio1* | A musical piece for guitar and VIVO that represents a typical example of pilot study. It includes a stochastic and user-defined computational score and a sound processing scenario, progressively refined through rehearsals. A camera detects the musician's physical motion in order to control the sound processing. |
| Velo*drone* | A cycling and musical event. The project aimed to extend the conventional concept of a competition and of a concert. Participants created the music by cycling. Physical motion related indirectly to sound generation as a type of cross-modal/extradiegetic mapping. |
| *Invisible Cities* | A short movie for prepared piano and VIVO that fits under the umbrella of remix cinema. Only motion in the video is mapped to the sound processing. Commonality of interaction between the musician and VIVO is achieved by |
| | monitoring changes in the motion threshold, as cues for the interplay (both intradiegetic and cross-modal/extradiegetic). |
| *Collective* | Sessions of improvisation comprising voice, trombone, guitar, double bass, drum kit, and dancer interacting with VIVO. The system is tested in interplay with multiple VIVO/performer instances and no user-defined score. The system provides multi-directionality to the informational feedback for/between the performers and denotes creative autonomy (see case study). |

*Gestural embedding*

In an acoustic instrument, the action-reaction cycle is at the basis of instrumentality and central to playing a musical instrument (Leman, 2008; Maes et al., 2014). Similarly, the principle of action/perception holds that when we excite the physical body of an acoustic instrument, we can see the direct relationship between our actions on it (action) and the sound that we hear (perception) in a process of identification-through-repetition (Emmerson, 2000). In the pilot studies on VIVO, the automatic sound generation exhibits **acousmatic** properties, as the sound processing forces the sources and causes of sound-making to become as 'remote or detached from known, directly experienced physical gesture and sounding sources' (Smalley, 1997, p. 112). Sound generation in VIVO allows designing action/perception feedback loops for music-making that are bond to a **causation mapping**. In this, a salient cue in the human interplayer is used to mould the automatic generation, which can be perceived both as autonomous and as retaining mutuality. The phenomenon is particularly relevant in an IMS, as a piece of information technology mediates the musical interaction, and consequently the sonic outcome and its source. This results in a complex and reiterated-but-changing mapping between action and sound. The scope of such a mediation surpasses physicality and perception, as projecting sound generation for the user to a cognitive dimension of musical expectancy.

In an IMS, the action-reaction mapping can therefore embody a sonification process where the quality of a gesture shapes the music. As mentioned in the previous section, motor knowledge can be embedded in VIVO through video-to-sound types of mapping. Such embodiment constitutes a musical goal-directedness for the human interplayer. VIVO works for the user as a means for the cultural embedding of gesture, which is typical in what is known as gestural surrogacy — the process of increasing **remoteness** (Smalley, 1997). Remoteness is a form of uncertainty that can be perceived in the causality between sound sources and sonic events. Here, remoteness is a factor enabling multidominance in the salient video-to-sound mappings. The generated sound here retains the imaginative property of an acousmatic sound. Such a multidominance-enabled mapping introduces novelty in the automatic generation.

In VIVO, the user's and audience's attribution of meaning to the automatically generated sound is dependent on causal action/perception relation suggested by the system. For the audience, this algorithmic generation is visible in the source from which the QoM is derived and computed (i.e., the video stream capturing a musician playing an instrument). For the user, the system's use of a variance-based threshold trigger allows the sound generation to change in correspondence to salient actions, ultimately increasing the coherence between the sound source and the acousmatic-like sound. Salience informs here the algorithmic generation, thereby preserving musical coherence in the interplay while also introducing gestural surrogacy.

*Broadening the action/perception feedback loop*

In the pilot studies, gestural surrogacy is established in VIVO when a directly mapped relation is formed between the user's gesture and the perception of the VIVO-generated outcome. Such action/perception feedback loop is therefore a critical component in delivering musical meaningfulness to the user. We have considered action/perception as a non-musical but experiential information for determining computer invention. A camera watching the user's body and the surrounding space (as first explored in *Studio1*) or a video file (as in *VIVOtube* and *Invisible Cities*) provided such information. In both cases, non-musical information drove the automatic generation and extended the user's agency in terms of gestural surrogacy and sense of coherence. This was achieved in different ways: (i) when instructions were sent to the machine for sound generation (as in the preparation of *VIVOtube*), (ii) through the processing of sound resulting from physical gestures on a musical instrument (as in all the pilots, with the exception of *VIVOtube* and *Velodrone*), (iii) through gestures on physical interfaces connected to software instruments (as in *Velodrone*), and (iv) through any gesture (e.g., dancing, as in *Collective*) or multimedia providing motion dynamics that could be mapped to a software instrument (as in *VIVOtube*, *Invisible Cities* and *Collective*). It should also be noted that there were instances wherein a performative gesture could not be mapped, for example when using a video file (as in *VIVOtube* and *Invisible Cities*), or when the interface was a physical device (as the bicycles in *Velodrone*).

In the pilot studies, VIVO was used to generate a simultaneous auditory feedback for each input information. The feedback generally referred to visual sequencing in a video file determining the QoM/SoA or a user/VIVO interaction in the physical space, or to the proprioception of users captured by a camera. We shall discuss further the theoretical nature of novelty and mutuality capacities implemented in VIVO as multidominance- and salience-based approaches, in terms of reflexivity and embodiment for the user.

*Enabling self-reflection mechanisms into VIVO*

Grounded cognition theories postulate that the brain intrinsically ties sensory information to the perceptual modality in which that information is perceived (Barsalou, 2008;

Pezzulo et al., 2013). According to such a view, both acoustic instruments and VIVO allow multimodal information to shift dynamically for the user, 'in reaction to the instrument and one's interaction with it' (Keebler et al., 2014). However, in contrast to acoustic instruments, VIVO is a piece of information technology that mediates (processes) and reflects (re-presents) the user's interactions.

In considering human cognition as embodied, VIVO was designed not only for facilitating the user's perception of system autonomy through stochastically-generated sonic constructs recognised as music by the user, but also for stimulating a subjective capacity for self-reflection. Self-reflection is thus implemented by design through interactions that imply rehearing, reproduction, and variation, across the visual and sonic domain. In such a self-reflection, both the perceived self and the perceiving self-mirror each other through musical constructs that embody the agent's activity. The specific meanings of the terms **emergence** and **reflection** refer here to processes of self-reflection in which musical constructs may embody agency. While the interpretation of an object is a process that operates multi-directionally and recursively in a semiotic/semiological loop, the process retrieves new information from new experiences and may potentially continue endlessly. In self-perception, both the perceived self and the perceiving self keep mirroring each other. Enclosed in a recursive loop of self-definition, the **I** is constituent to the same self. However, it extends over the boundaries of individual reflection; self-consciousness surpasses self-reflection, as the **I** is the result of an interpretative process which culture incorporates. VIVO affords the users with a complementary way to extend their inner body knowledge in order to experience self-reflection in music. A mapping between motion tracking algorithms, sound generation and the user interface is established. The considerations drawn from the pilot studies in the present article suggest that self-reflection may be a property of the network of interaction that is established between VIVO and the user. Interestingly, some backing to this can be found in recent research on consciousness as a state of matter, rather than as an emerging property (Tegmark, 2015). Similarly, the reflexivity of the interaction and its character of multidominance may constitute a state of the user/system network where the capacity for meaningfulness does not emerge from an evolving process of interaction between human and software agents but rather is inherent to the system properties since the system design. In the next section, we evaluate our working hypothesis stating that implementations incorporating a bare minimum of novelty (multidominance) and mutuality (salience) capacities is a sufficient condition for IMSs, as well as VIVO, to exhibit creative autonomy.

## 4. Case study: excerpt from Collective

Figure 4 (also see Supplemental Material) depicts a transcription of the audio recording from Collective (Table 1), illustrating a free improvisation between a trombone player and VIVO. The transcription was generated automatically from the recording via the automatic music transcription software Melodyne 4 (Celemony Software, 2017) using standard settings for polyphonic music. The transcription was adjusted manually in the engraving process for both the trombone and the VIVO parts of the score to reflect the actual playing. In the VIVO part, only salient cues are engraved, in contrast to greying out parts where timbre is considered predominant over pitch.

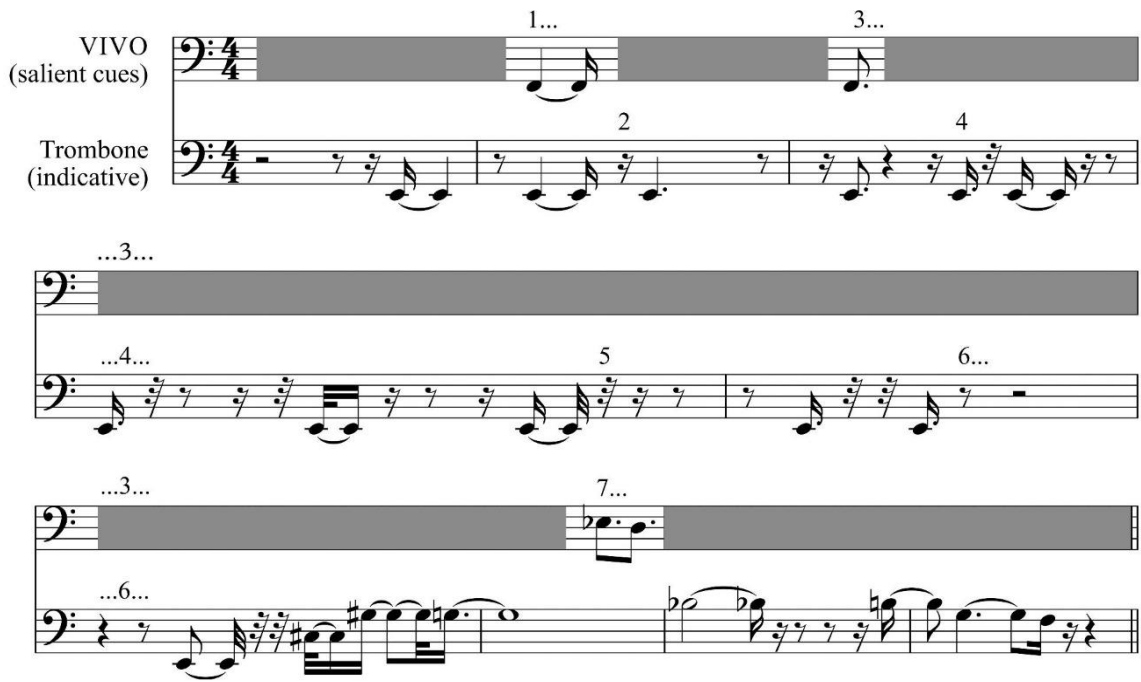

Figure 4. Excerpt of score (automatic transcription) from Collective.

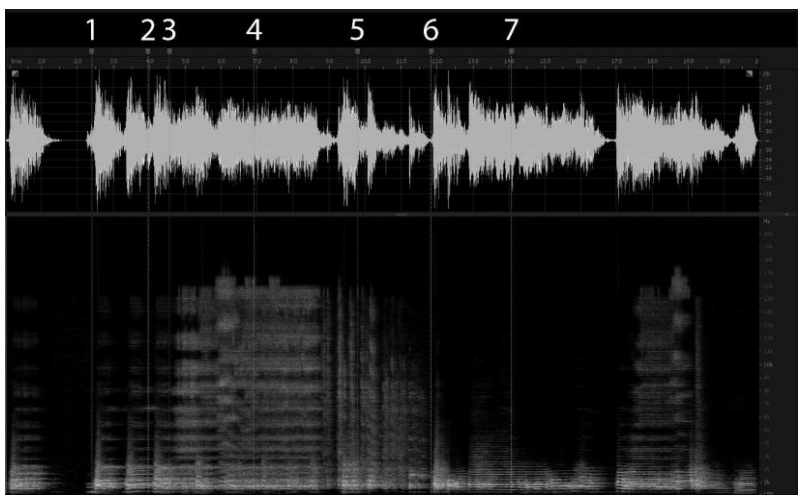

Figure 5. Excerpt of spectrogram (left + right) from Collective.

Here, VIVO interplays with a trombone player (Figure 4-1 and Figure 5-1) and the musician's response results in the activation of the variance-based threshold trigger (4-2 and 5-2). The musician recognises the consequent sound generation as an opportunity for action to achieve a meaningful interplay with the system. The musician also recognises that can retain only a partial control over the system — the system is autonomous. This is verified as the musician listens to VIVO (4-3 and 5-3) and then shapes his own playing accordingly (4-4 and 5-4) thereby activating again the threshold trigger again (4-5 and 5-5). Notably, the musician's achievement of musical phrasing after the first trigger (4-6 and 5-6) confirms the intentionality of this second trigger. VIVO's interplay initiation depends on the musician's playing but although the system denotes a certain level of autonomy, the interplay remains coherent. Furthermore, the musician does not attain musical coherence casually, for example, by independently adding his own playing to the computer generation. Instead, the musician achieves musical phrasing in what we can describe as a form of **reflexive multidominance**. By mapping the QoM to audio parameters and the SoA to the sound generation, the musician influences the system, while in turn the system influences the musician by providing sonic responses favouring a reflexive interplay (Figures 4-6 and 4-7, 5-6 and 5-7).

In the present case study, VIVO works for the musician both as an instrument and as an autonomous player. This mode of operation echoes Rowe's definition of an IMS at both extremes of the continuum proposed in that definition, dynamically. As an instrument, the system extends the musician's capacity for music-making through an embodiment of the control for sound generation; the system is here an extension of the acoustic instrument controlled through physical gestures. As an autonomous player, the mapping serves as reference for the automatic generation through a stochastic score embedding the audio plug-ins loaded in the dynamic host and using the QoM and SoA from the musician's gestures on the acoustic instrument. Because the musician engages in reflexive multidominance through the interplay, the present study also suggests that system autonomy is here a property shared with the user.

## Conclusions

In the present article, autonomy in IMSs has been discussed as a pivotal capacity for self-determination, yet not sufficient for a machine to be autonomously creative as a human agent. Creative autonomy has been investigated as a compound property that incorporates the capacity to introduce novelty in the generation and retain mutuality to the content of interaction. We have described the implementation of this capacities in terms of salience and multidominance as the bare minimum of properties implemented for creative autonomy to be identified in the system. We have presented VIVO, an IMS which incorporates salience and multidominance via reflexive interactions with the system. We have provided details of a case study presenting a musical evidence. In this the automatic generation denotes both mutuality and novelty, and the musician's interplay describes a sound production that is aware of the system's autonomy. In future studies, we will draw specific comparisons to other systems, as well as investigate further the proposed concept of reflexive multidominance as a property for meaningfulness of the system.

**Funding**

The present research has received funding from the European Union's Horizon 2020 research and innovation programme under the Marie Sklodowska-Curie grant agreement

No 659434.